\documentclass[twocolumn,numberedappendix,twocolappendix,appendixfloats]{openjournal_dhayaa}

\usepackage{amsmath}
\usepackage{fontawesome5}
\usepackage{color}
\usepackage{xcolor}

\usepackage{natbib}
\setcitestyle{aysep={}}

\usepackage[colorlinks=true
  ,urlcolor=blue
  ,anchorcolor=blue
  ,citecolor=blue
  ,filecolor=blue
  ,linkcolor=blue
  ,menucolor=blue
  ,linktocpage=true
  ,pdfproducer=medialab
  ,pdfa=true
]{hyperref}

\newcommand{\Mtwohc}{M_{\rm 200c}}

\newcommand{\Rvir}{R_{\rm vir}}
\newcommand{\Mvir}{M_{\rm vir}}

\newcommand{\cvir}{c_{\rm vir}}

\newcommand{\eg}{{\sl e.g.}, }     
\newcommand{\dex}{{\,\,\rm dex}}

\newcommand{\msol}{\ensuremath{\, {\rm M}_\odot}}    
\newcommand{\msun}{\ensuremath{\, {\rm M}_\odot}} 
         
\newcommand{\mpc}{\ensuremath{\, {\rm Mpc}}}

\definecolor{orcidlogocol}{HTML}{A6CE39}
\definecolor{purple}{RGB}{128, 0, 128}

\newcommand{\OrcidID}[1]{ \href[urlcolor = red]{https://orcid.org/#1}{\textcolor{lightgray}{\faOrcid}}}
\newcommand{\OrcidIDName}[2]{\href{https://orcid.org/#1}{#2}}

\defcitealias{Anbajagane2022BaryImprint}{A22}
\defcitealias{Shao2023Baryons}{S23}
\defcitealias{Singh2020CosmologyWithMORs}{S20}

\newcommand{\Shao}{Shao2023Baryons}
\newcommand{\Anba}{Anbajagane2022BaryImprint}

\usepackage[T1]{fontenc}
\usepackage{ae,aecompl}

\usepackage{array}
\newcolumntype{M}[1]{>{\raggedright\arraybackslash}m{#1}}
\newcolumntype{C}[1]{>{\centering\arraybackslash}m{#1}}

\newcommand{\param}[1]{%
  \ifcase#1\relax
  \or \texttt{Omega0}%
  \or \texttt{Sigma8}%
  \or \texttt{WindEnergyIn51erg}%
  \or \texttt{RadioFeedbackFactor}%
  \or \texttt{VariableWindVelFactor}%
  \or \texttt{RadioFeedbackReiorientationFactor}%
  \or \texttt{OmegaBaryon}%
  \or \texttt{HubbleParam}%
  \or \texttt{n\_s}%
  \or \texttt{MaxSfrTimescale}%
  \or \texttt{FactorForSofterEQS}%
  \or \texttt{IMFslope}%
  \or \texttt{SNII\_MinMass\_Msun}%
  \or \texttt{ThermalWindFraction}%
  \or \texttt{VariableWindSpecMomentum}%
  \or \texttt{WindFreeTravelDensFac}%
  \or \texttt{MinWindVel}%
  \or \texttt{WindEnergyReductionFactor}%
  \or \texttt{WindEnergyReductionMetallicity}%
  \or \texttt{WindEnergyReductionExponent}%
  \or \texttt{WindDumpFactor}%
  \or \texttt{SeedBlackHoleMass}%
  \or \texttt{BlackHoleAccretionFactor}%
  \or \texttt{BlackHoleEddingtonFactor}%
  \or \texttt{BlackHoleFeedbackFactor}%
  \or \texttt{BlackHoleRadiativeEfficiency}%
  \or \texttt{QuasarThreshold}%
  \or \texttt{QuasarThresholdPower}%
  \else
    \texttt{UnknownParam}%
  \fi
}

\begin{document}

\title{Baryonic Imprints on DM Halos: the concentration--mass relation and its dependence on 28 model parameters in the CAMELS suite}

\author{\OrcidIDName{0009-0007-4233-4470}{Cami Delgado}$^{1, \star}$}
\author{\OrcidIDName{0000-0003-3312-909X}{Dhayaa Anbajagane}$^{1, 2, \star}$}
\email{$^{\star}$ dhayaa@uchicago.edu}
\affiliation{$^{1}$ Department of Astronomy and Astrophysics, University of Chicago, Chicago, IL 60637, USA}
\affiliation{$^{3}$ Kavli Institute for Cosmological Physics, University of Chicago, Chicago, IL 60637, USA\\}

\begin{abstract}
The evolution of baryons in halos, as well as their gravitational impact on the total-matter phase space, has a rich phenomenology and is vital in answering many questions in both cosmology and astrophysics. We use the \textsc{Camels} simulation suite to quantify the impact of twenty-three different galaxy formation parameters/processes (as well as five cosmological parameters) on the concentration–mass relation, $\cvir - \Mvir$, which characterizes the radial density distribution in halos. We construct a simulation-informed nonlinear model for concentration as a function of halo mass, redshift, and twenty-eight cosmological/astrophysical parameters. This is done using 2048 simulations of the IllustrisTNG galaxy formation model from the \textsc{Camels} suite. We extract the imprints of galaxy formation across a wide range in mass $\Mvir \in [10^{11}, 10^{14.5}]$ $M_{\odot}/h$ and in redshift $z \in  [0, 3]$, finding clear mass- and redshift-dependent features in each of the different feedback mechanisms. From our measurements, we construct a three-component phenomenological model that can include astrophysical imprints in the concentration--mass relation in a fast and simple manner.
\end{abstract}



\section{Introduction}

Halos are key building blocks of the structure observed in our Universe. Through the so-called ``halo model'' approach \citep{Cooray2002HaloModel,Asgari2023HaloModel}, one combines three principal ingredients --- the abundance of halos, their spatial clustering, and the internal distribution of quantities like density, pressure, temperature, etc. --- to predict a wide variety of cosmological observables. Applications range from matter and galaxy density fields to electron-pressure and X-ray photon-count fields \citep[\eg][]{Asgari2023HaloModel,Mead2020HMx,Ferreira2024XrayShear,Anbajagane2026PrimordialNonlinear}, and have enabled joint constraints on both cosmology and baryonic astrophysics \citep[\eg][]{Troster2022KidstSZ,Dvornik2023GGLKiDs,Pandey2023Baryons, Anbajagane2023Shocks, Siegel:2026:baryons}. Consequently, considerable effort has gone into characterizing each of these three ingredients, with major advances over the past two decades driven by increasingly high-resolution simulations of structure formation \citep[e.g.][]{Klypin2016Concentrations,Vogelsberger2014Illustris,Schaye2015EAGLE,McCarthy2017BAHAMAS,Pillepich2018Methods}. Of these ingredients, the halo internal structure is arguably the most challenging to model precisely given it is sensitive to a particularly diverse phenomenology --- including density-profile shape and concentration, assembly history, triaxiality, substructure, and baryonic modification --- that are challenging to capture \citep[e.g.][]{Navarro1997NFWProfile,Bullock2001Concentrations,Klypin2016Concentrations,Ludlow2016Concentration,Diemer2019concentrations,Schaller2015ConcentrationBaryonImprints,Anbajagane2022BaryImprint}.

This phenomenology arises as astrophysical processes can dramatically change the internal structure of halos, and these processes exhibit a rich phenomenology. Significant efforts have been undertaken to characterize these astrophysical effects, particularly on the matter density profile of halos. \citet{Duffy2010BaryonDmProfileDensity} simulated the growth of structure alongside a variety of different astrophysical prescriptions, and found the impact of star formation and metal cooling had notable impacts on the resulting density distribution. \citep{Henson2017BaryonEffectsBM} used a high-resolution simulation suite, including a number of zoom-in simulations, to extract precision estimates of internal structure in galaxy clusters. Other works \citep{Ragagnin2019HaloConcentration, Beltz-Mohrmann2021BaryonImpactTNG} have similarly characterized this impact using different simulations of galaxy formation and structure. \citet[][henceforth \citetalias{\Anba}]{Anbajagane2022BaryImprint} compared the concentration--mass relation of halos from dwarf-galaxy  mass scales to galaxy-cluster mass scales using the IllustrisTNG simulations \citep{Springel2018FirstClustering, Nelson2018FirstBimodality, Naiman2018FirstEuropium, Marinacci2018FirstFields, Pillepich2018FirstGalaxies}, and identified non-monotonic, mass-dependent variations that are directly linked to the impact of stellar feedback, gas cooling, and active galactic nuclei (AGN) feedback.

The above works primarily focused on fixed galaxy-formation models, i.e., models in which the equations governing unresolved astrophysical processes (and the associated free parameters) were set after calibrating against a selected set of observations \citep[\eg][]{Vogelsberger:2013:TNGModel,Schaye2015EAGLE,Crain2015EAGLECalibration,McCarthy2017BAHAMAS,Pillepich2018Methods}. The precise set of calibration observables varies between simulation suites, but commonly include the galaxy stellar mass function, galaxy sizes, black-hole--galaxy scaling relations, the cosmic star-formation history, and the hot-gas fractions of galaxy groups and clusters. The \textsc{Camels} collaboration \citep{Villaescusa-Navarro2021CAMELS} produced a suite of small-box simulations where these free parameters are not held fixed but are instead systematically varied. \citet[][henceforth \citetalias{\Shao}]{Shao2023Baryons} used earlier versions of this suite to show the specific, mass-dependent impact of four astrophysical parameters (pertaining to supernovae and AGN feedback), and two cosmological parameters, on the concentration--mass relation of halos. Halo concentrations are apt summarizes of the radial density profile of a halo \citep{Navarro1997NFWProfile, Navarro2010EinastoVsNFW}, and are therefore simpler to work with. The individual simulations span a volume of $V = (25 {\rm Mpc/h})^3$, but one can use the combination of the $\mathcal{O}(10^3)$ simulations provided by \textsc{Camels} to build a ``uber'' sample with a larger effective volume. Using such an approach, \citetalias{\Shao} built a non-parametric, quasi-linear model for the concentration--mass relation and shed light on the relationship between concentration and specific simulation hyperparameters related to supernovae (SNe) and AGN feedback.

In this work, we use a new set of 2048 simulations from \textsc{Camels} to extend the model of \citetalias{\Shao} --- which tested the dependence of the halo concentration on 6 model parameters --- to now test the dependence of halo concentration on 28 parameters. The parameter set spans a much wider variety of astrophysical and cosmological contributions, and presents a more complete characterization of the dependence of halo concentration on these processes. We focus on the \textsc{IllustrisTNG} model presented in \textsc{Camels}, and use the same non-parametric modeling approach of \citetalias{\Shao}, which we will detail further below. The broad set of parameters being varied allows us to more completely catalog the connection between halo concentration and various astrophysical processes, and to define subgroups of parameters based on similarities in their behavior. This larger parameter set will also let us better highlight potential non-linear couplings between processes, following earlier work in \citetalias{\Shao}.

Our paper is organized as follows: in \S \ref{sec:DataMethods} we describe the \textsc{Camels} simulation data, paying particular attention to the halo concentration measurements, and describe the kernel-localized linear regression (\textsc{Kllr}) method used to analyze the data. In \S \ref{sec:results} we extract the full dependence of the concentration--mass relations on the five cosmology parameters and the twenty-three astrophysical parameters, and then build an effective model that can capture the behavior of the latter parameters. We conclude in \S \ref{sec:conclusions}.

\section{Data and Methods}\label{sec:DataMethods}

\subsection{The \textsc{Camels} simulations}\label{sec:Camels}
The Cosmology and Astrophysics with MachinE Learning Simulations \citep[\textsc{Camels,}][]{Villaescusa-Navarro2021CAMELS}, is a simulation suite of $V = (25 \mpc/h)^3$ boxes, each with varying cosmological and astrophysical parameters. We use the SB28 run, which varies 28 different parameters with their values sampled through a Sobol grid sequence. The SB28 runs exist for different galaxy formation models, of which we focus on the \textsc{IllustrisTNG}, or TNG, model for brevity and given the halo concentration in TNG models have been extensively studied \citep[\eg][]{Beltz-Mohrmann2021BaryonImpactTNG, Anbajagane2022BaryImprint, Shao2023Baryons} and so, has multiple reference points that can be taken advantage of in our work. The full set of 28 parameters being varied is listed in Table \ref{tab:params}. We include the ranges the parameters are varied within, references to the exact differential equations they modify, and a brief interpretation of the physical parameter itself. The parameter choices were introduced in \citet[][see their Appendix A.1]{Ni2023Camels28}, and we mimic their parameter ordering and naming schemes in our presentation here for better clarity.

The \textsc{Camels} simulations were run with $256^3$ dark matter particles of mass $6.5\times 10^7 (\Omega_{\rm m} - \Omega_{\rm b})/0.251 \msol/h$, and then $256^3$ gas resolution elements with an initial mass of $1.27 \times 10^7 \msol/h$. The suite is specialized for training machine learning algorithms but the dataset can also be used for non-parameteric studies of astrophysics and cosmology, such as the one pursued in this work. Each of the 2048 SB28 simulations is run with different initial conditions, i.e. different random seeds. The \textsc{Camels} simulation suites contains multiple galaxy formation models. As mentioned previously, we use the runs from the TNG galaxy formation model. The choices of this model are described in detail in \citet{Weinberger2017Methods, Pillepich2018Methods}. The resolution level of the \textsc{Camels} simulations roughly corresponds to the TNG100-2 run \citep{Nelson2019TNGPublicData}. While \textsc{Camels} currently also contains a SB28 suite for the SIMBA model, we focus only on the TNG suite here for simplicity. Table \ref{tab:params} provides an overview of the different parameters and their interpretations, including explicit equations that introduce each astrophysical parameter.

We utilize the \textsc{Rockstar} halo catalogs \citep{Behroozi2013Rockstar}, which identify halos as clusters in position-velocity space, from each simulation. Halos are defined with their spherical overdensity radii, $\Rvir$, defined as the radius within which the mean density in the halo is $\rho = \Delta_{\rm vir}(z) \rho_{\rm c}(z)$. The contrast $\Delta_{\rm vir}$ is roughly around $180$ at $z = 0$ in a fiducial cosmology \citep{Bryan1998vir}. The halo mass $\Mvir$ is defined as the mass contained within a sphere of radius $\Rvir$. The concentration in \textsc{Rockstar} is obtained by fitting total matter density profile with the functional form,
\begin{equation}\label{eqn:NFW}
    \rho(r) = \frac{\rho_s}{x (1 + x)^2}
\end{equation}
where $\rho_s$ is the scale density, and $x \equiv r/r_s$ is the radius normalized by the scale radius, $r_s$. The concentration is then defined as, $\cvir = \Rvir / r_s$. 

We emphasize that \textsc{Rockstar} fits Equation \eqref{eqn:NFW} to the total matter distribution, rather than just to the dark matter. The resulting $\cvir$ estimates will incurr some error due to model mis-specification, as the true total-matter distribution will not follow the NFW profile. However, the key requirement in this work is a simple metric that summarizes the total matter distribution of the halo, and the concentration measured this way is adequate in doing so as it is estimated using this full matter distribution. \citetalias{\Shao} found their concentration--mass relation in \textsc{Camels} followed the trends observed in \citetalias{\Anba}, where the former (latter) work used $\cvir$ from the total matter (dark matter-only) distribution.

\begin{table*}
    \centering
    \begin{tabular}{|M{0.23\linewidth}|C{0.12\linewidth}|C{0.06\linewidth}|C{0.11\linewidth}|M{0.35\linewidth}|}
        \hline
        \multicolumn{1}{|c|}{\textit{Param Number / Name}} &
        \multicolumn{1}{c|}{\textit{Range}} &
        \multicolumn{1}{c|}{\textit{Type}} &
        \multicolumn{1}{c|}{\textit{Equation}} &
        \multicolumn{1}{c|}{\textit{Description}} \\[5pt]
        \hline
        1: Omega0 & [0.1, 0.5] & Cosmo & N/A & The fractional matter energy density, $\Omega_{\rm m}$ \\
        \hline
        2: Sigma8 & [0.6, 1] & Cosmo & N/A & The root-mean square fluctuations of the $z = 0$ density field, smoothed on apertures of $8 \mpc/h$.\\
        \hline
        3: WindEnergyIn51erg & [0.9, 14.4] & SNe & P18 - 3  & The energy scale of SNe explosions, in units of $10^{51}$ ergs.\\
        \hline
        4: RadioFeedbackFactor & [0.25, 4.0] & AGN & W17 - 8 & The energy per AGN event, in the radio mode \\
        \hline
        5: VariableWindVelFactor  & [3.7, 14.8] & SNe & P18 - 1 & Galactic wind speed\\
        \hline
        6: RadioFeedbackReiorientationFactor & [10, 40] & AGN & W17 - 13 & The rate of AGN events in the low-accretion mode\\
        \hline
        7: OmegaBaryon & [0.029, 0.069] & Cosmo & N/A & The fractional baryon enery density, $\Omega_{\rm b}$\\
        \hline
        8: HubbleParam & [0.4711, 0.8711] & Cosmo & N/A & The Hubble expansion parameter, $h = H_0 / 100$\\
        \hline
        9: n\_s & [0.7624, 1.1624] & Cosmo & N/A & The spectral index of the primordial power spectrum \\
        \hline
        10: MaxSfrTimescale [Gyr] & [1.135, 4.54] & SNe & S\&H03 - 21 & The timescale for star-formation when a gas cell reaches the requisite density threshold for star-formation\\
        \hline
        11: FactorForSofterEQS & [0.1, 0.9] & SNe & S05 & A factor interpolating between effective EoS for star-forming gas and isothermal EoS for $T = 10^4 K$ gas\\
        \hline
        12: IMFslope & [$-2.8$, $-1.8$] & SNe & C03 & Power-law index of the stellar initial mass function above $1 \msun$. \\
        \hline
        13: SNII\_MinMass\_Msun [$M_{\odot}$] & [4, 12] & SNe & V13 - 11 & The lowest mass star that produces a SNe explosion\\
        \hline
        14: ThermalWindFraction & [0.025, 0.4] & SNe & P18 - 2,4,5 & Fraction of galactic winds injected thermally instead of kinetically\\
        \hline
        15: VariableWindSpecMomentum [km/s] & [0, 4000] & SNe & V13 - 15 & The specific momentum of galactic winds (in practice, changes the mass-loading factor but not the wind speed)\\
        \hline
        16: WindFreeTravelDensFac & [0.005, 0.5] & SNe & & Gas density at which collisionless galactic winds couple back to the hydrodynamic fluid\\
        \hline
        17: MinWindVel [km/s] & [150, 550] & SNe & P18 - 1 & Minimum galactic wind velocity\\
        \hline
        18: WindEnergyReductionFactor & [0.0625, 1.0] & SNe & P18 - 3,4 & Galactic wind energy at high-metallicity relative to lower metallicity\\
        \hline
        19: WindEnergyReductionMetallicity & [0.5, 8] $\times 10^{-3}$ & SNe & P18 - 3,4 & Metallicity value that sets the transition from high to low metallicity \\
        \hline
        20: WindEnergyReductionExponent & [1, 3] & SNe & P18 - 3 & Sharpness of the metallicity-dependent transition in galactic wind speeds \\
        \hline
        21: WindDumpFactor & [0.2, 1.0] & SNe & V13 - 20 & Fraction of metals in star-forming cells injected into neighboring cells and not ejected via winds\\
        \hline
        22: SeedBlackHoleMass [$M_{\odot}$] & [2.5, 25] $\times 10^5$ & AGN & V13 & Mass of supermassive black hole seed\\ 
        \hline
        23: BlackHoleAccretionFactor & [0.25, 4] & AGN & W17 - 2 & Scales the Bondi rate of accretion onto a black hole\\
        \hline
        24: BlackHoleEddingtonFactor & [0.1, 10] & AGN & W17 - 3 & Scales the Eddington rate of accretion onto a black hole \\ 
        \hline
        25: BlackHoleFeedbackFactor & [0.025, 0.4] & AGN & W17 - 7 & Energy of AGN feedback in quasar mode\\ 
        \hline
        26: BlackHoleRadiativeEfficiency & [0.05, 0.8] & AGN & W17 - 7,9 & The fraction of accretion rest-mass that is radiated away as energy \\ 
        \hline
        27: QuasarThreshold & 63 $\times [10^{-6}, 10^{-2}]$ & AGN & W17 - 5 & Eddington ratio threshold that distinguishes low-accretion and high-accretion systems \\ 
        \hline
        28: QuasarThresholdPower & [0, 4] & AGN & W17 - 5 & Power-law slope of the Eddington ratio threshold with black hole mass\\ 
        \hline
        
    \end{tabular}
    \caption{The 28 different parameters varied in the \textsc{Camels} SB28 suite, alongside the parameter ranges being sampled by the suite, and the specific equations each astrophysical parameter corresponds to as listed in the following papers: P18 \citep{Pillepich2018FirstGalaxies}, W17 \citep{Weinberger2017Methods}, S\&H03 \citep{Springel:2003:Cosmo}, and V13 \citep{Vogelsberger:2013:TNGModel}. The last column provides a heuristic description for each parameter. See Appendix A1 in \cite{Ni2023Camels28} for more details.}
    \label{tab:params}
\end{table*}

\subsection{Kernel-localized linear regression}\label{sec:KLLR}

Halo-property scaling relations are well fit using a simple power law, i.e. a log-linear relation \citep[\eg][]{Evrard2008VirialScaling, Allen2011CosmoClusterReview, Kravtsov2012ClusterFormation, Kravtsov2018SMHMRelation, Lim2021GasMass, Lee2022rSZ}. Others can be well fit using extended parametric models such as a broken power-law model with two power-law indices instead of one \citep[][see their Figure 2 and references therein]{Wechsler2018GalaxyHaloConnection}. Such parameterizations need not capture all the non-linearity in a scaling relation. This is especially true in hydrodynamical simulations, where galaxy formation introduces many unique mass/time-scales \citep[\eg][]{Crain2015EAGLECalibration, Pillepich2018Methods, Weinberger2017Methods, Vogelsberger2020Hydro} and thereby causes scaling relations for most halo properties to deviate from a ``self-similar'', power-law behavior \citepalias{\Anba}. Given the surplus of available simulation data, we can utilize data-driven models that capture these non-linearities scalings with halo mass without having to assume fixed functional forms.

One approach to building model-independent representations of scaling relations is to use machine learning (ML) techniques, such as random forest, neural networks, and more \citep[\eg][]{Machado2020GasShapesSHAP, Stiskalek2022ML}. Such methods are also more difficult to interpret, and thereby more difficult to extract physical explanations from. Though there are approaches that can alleviate the interpretability issue \citep[\eg][]{Ntampaka2019XrayClustersML, Machado2020GasShapesSHAP, \Anba}. An alternative then is to continue with well-known, well-understood linear models but extend them to capture some or most of the non-linearity introduced by astrophysics.

Kernel-localized linear regression \citep*[\textsc{Kllr},][]{Farahi2022KLLR} is an extension that allows such flexibility. The input data (i.e. halos) are weighted according to a kernel, which is chosen to be a Gaussian kernel in $\log_{10}\Mvir$, and a linear regression is fit to the kernel-weighted data. As the kernel is shifted in the log-mass direction, the method measures the non-linear scaling relation using multiple, connected linear pieces. \textsc{Kllr} has had extensive use in extracting nonlinearity in different halo property--mass scaling relations \citep[][\citetalias{\Anba}, \citetalias{\Shao}, \citealp{Shao:2024:Baryons}]{Farahi2018BAHAMAS, Anbajagane2020StellarStatistics, Anbajagane2022GalVelBias}. In this work, we continue on the efforts of \citetalias{\Shao} and use \textsc{Kllr} to study a multivariate scaling relation that is still modeled as being locally linear in halo mass. 

The locally linear, multivariate scaling relation of a halo property is written in \textsc{Kllr} as,
\begin{align}\label{eqn:Kllr}
    \log_{10}\cvir = &\,\, \pi(\Mvir, z) + \alpha_M(\Mvir, z) \log_{10}\Mvir\nonumber\\
    & + \sum_X\alpha_X(\Mvir, z) f_X(X),
\end{align}
where $\pi$ is the intercept, $\alpha_M$ is the slope with halo mass, $\alpha_X$ is the slope with the other twenty-eight parameters. The slopes and intercepts are all functions of halo mass and redshift. Note that the dependence of $\cvir$ on the cosmological/astrophysical parameters is also a function of mass. This allows us to capture mass-dependent changes to the relationship between the halo concentration and a given parameter. The function, $f_X(X)$ is either $f_X(X) = X$ or $f_X(X) = \log_{10}X$ depending on whether the SB28 suite sampled over $X$ or $\log_{10}X$.

We compute a \textsc{Kllr} model for a series of six snapshots in $\textsc{Camels}$, using the uber sample at each redshift between $0 < z < 3$. In practice, this means we compute Equation \eqref{eqn:Kllr} independently for the halo sample from each redshift. Our results in this work will focus mostly on the model at $z = 0$ but we present and discuss the redshift evolution of the model as well. We compute the scaling relations for every $0.1 \dex$ in mass, with the minimum mass set at $10^{11} \msun/h$ and a maximum mass set by the 20th most massive halo of the uber sample. The choice of upper limit excludes mass regimes where the model solutions are imprecise given the small sample size. The Gaussian kernel in $\log_{10}\Mvir$ is defined by $\sigma = 0.2 \dex$. Uncertainties on all the \textsc{Kllr} parameters of Equation \eqref{eqn:Kllr} are obtained by estimating regression coefficients for 100 bootstrap realizations of the data. Thus, we have 100 versions of Equation \eqref{eqn:Kllr} that then quantify the uncertainty in the predictions.

Other techniques, such as Gaussian processes, are also valuable in capturing the non-linearity of a relation in a non-parametric way with minima assumptions. For example, \textsc{Kllr} and Gaussian processes have been previously used to extract the same non-linear relations in halo quantities \citep[\eg][]{Farahi2021PoPE, Farahi2022ProfileCorr}. In this work, we continue using \textsc{Kllr} given it has already been used in characterizing halo scaling relations for many different properties.

\section{Results} \label{sec:results}

We now present our results for the dependence of the concentration--mass relation on cosmology (Section \ref{sec:results:cosmo}) and astrophysics (Section \ref{sec:results:SNeastro} and \ref{sec:results:AGNastro}). While we separate the two in our discussions, their impact can be non-linearly coupled and so we will note these where relevant. We then introduce a simple principle-component model of these baryon imprints, built using our measurements, in Section \ref{sec:results:PCA}. We briefly discuss the high-redshift signatures of these effects in Section \ref{sec:results:zdep}.

Almost all of our discussions will focus on the slopes between the halo concentration and the various input parameters. In particular, we show the modified quantity, $\hat{\alpha}$. For input quantities that are linear, and not logarithmic, the corresponding slopes $\alpha$ are coefficients with dimensions.\footnote{We have the relation $\ln \cvir = \alpha_x x$. Given the right-hand side must be dimensionless, $\alpha_x$ has units. For the relation, $\ln \cvir = \alpha_x \ln x$, the slope is already dimensionless.} This complicates the interpretation of their amplitudes, as latter can be arbitrarily scaled by changing the units of the chosen dimensions. Such an ambiguity is not present in the slopes corresponding to logarithmic inputs. In order to simplify comparisons between the slopes with logarithmic/linear quantities, we perform the transformation, $\hat{\alpha}_x \equiv \alpha_x x_{\rm fid}$ for any linear quantity as it removes the units from the slope. Here, $x_{\rm fid}$ is the fiducial value of a given parameter.\footnote{For the \param{15} parameter in Table \ref{tab:params}, the fiducial value is 0, which would simply make $\hat{\alpha} = 0$ under our transformation. Instead, we use 1000 $\rm km/s$ as its fiducial value.} Under this transformation, $\hat{\alpha}$ is a coefficient that converts fractional shifts in a parameter $x$ to fractional shifts in $\cvir$.  Thus, the slopes we discuss below can be viewed as the derivative of $\ln \cvir$ with a given input parameter. It is not advantageous to forcefully use the logarithmic transform of all input params as some parameters have $x \leq 0$, and are also sampled uniformly in linear (not log) space.

\subsection{Dependence on cosmology} \label{sec:results:cosmo}

\begin{figure*}
    \centering
    \includegraphics[width=2.1\columnwidth]{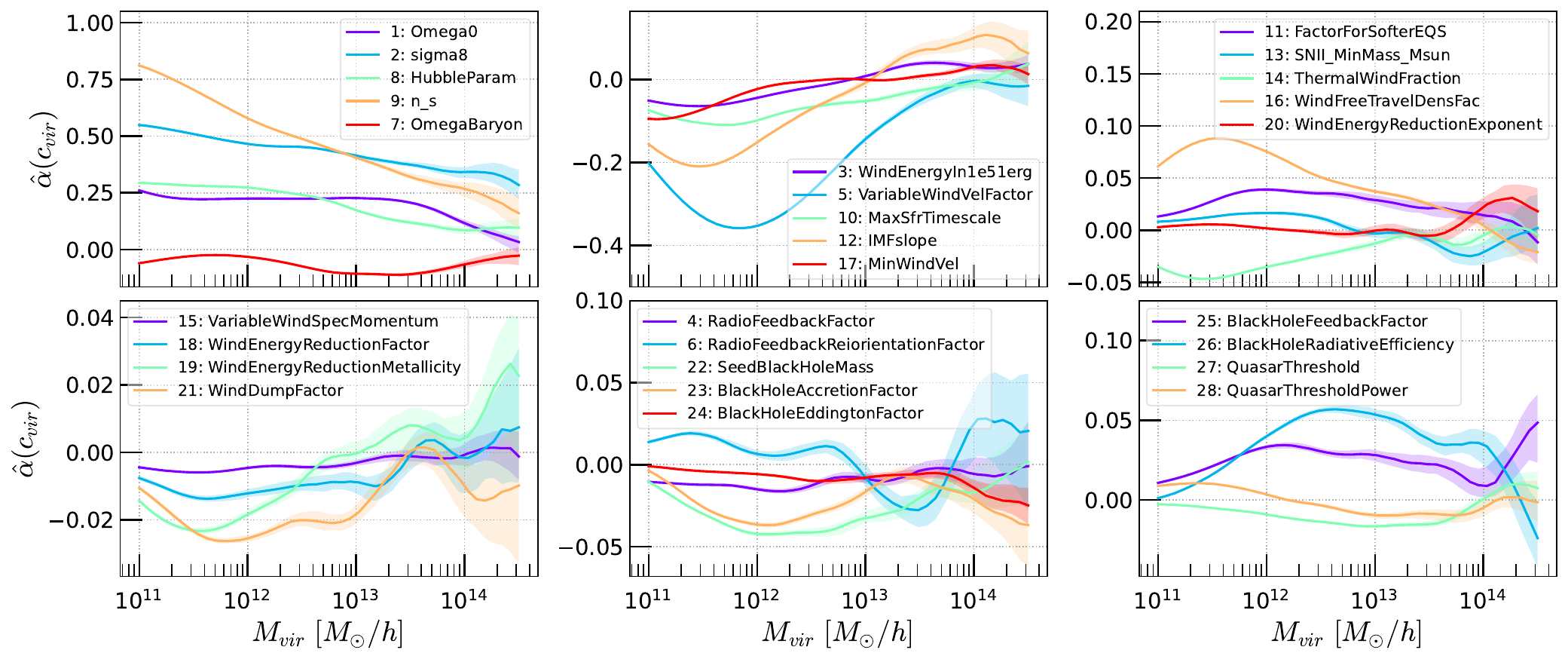}
    \caption{The slopes of concentration $\cvir$ as a function of halo mass $\Mvir$ and the 28 cosmological and astrophysical parameters. To enable better visibility, we organize parameters into different subsets based on the amplitudes of their slopes, and the category of parameters they fall into. The upper left panel contains the cosmology parameters, while the next three panels (upper middle, upper right, and lower left) contain SNe parameters, and the remaining ones (lower middle and lower right) are AGN parameters. There is a wide range of behaviors, which we discuss further in Sections \ref{sec:results:cosmo}, \ref{sec:results:SNeastro}, and \ref{sec:results:AGNastro}.}
    \label{fig:slopes}
\end{figure*}

\begin{figure*}
    \centering
    \includegraphics[width=2\columnwidth]{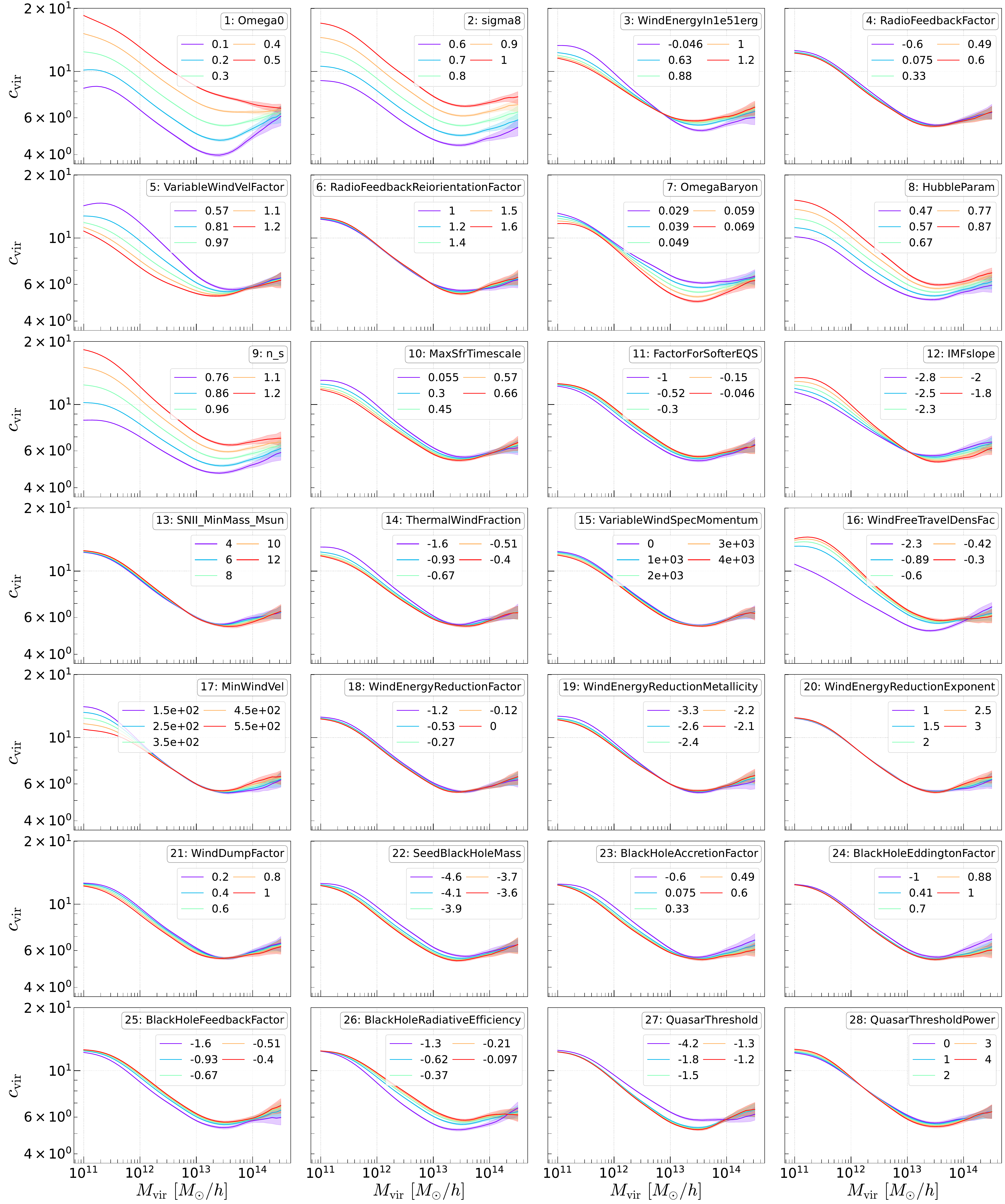}
    \caption{Predictions for the concentration--mass relation at $z = 0$, according to the KLLR model. Each panel varies one of the 28 parameters, with the values denoted in the legend. The variations are determined by the slopes, and so the above is a different demonstration of the results in Figure \ref{fig:slopes}.}
    \label{fig:cM_relation_variance}
\end{figure*}

We extract the slope of 28 different parameters and present them in Figure \ref{fig:slopes}, while the impact on the $\cvir$-$\Mvir$ relation is shown in Figure \ref{fig:cM_relation_variance}. We group the slopes in the former plot roughly by the magnitude of their slopes, in order to more easily pick out their behaviors, and by whether they correspond to cosmology, or to SNe or AGN astrophysics. The dependence on cosmology (upper left panel) has the largest amplitude, as expected. For example, $\Omega_{\rm m}$ and $h$ both have a similar impact on the halo concentration. This can be expected as they both directly scale the matter density, $\rho_{\rm m}$\footnote{The Hubble constant scales the critical density, $h^2 \propto \rho_c$, which informs the matter density, $\rho_{\rm m} = \Omega_{\rm m} \rho_c$.}, as well as the growth rate of density perturbations \citep{Dodelson2003ModernCosmology, Linder2005GrowthHistory}. The dependence on $\sigma_8$ follows previous works \citepalias[][]{\Shao} and can be related to the formation history of the halo (also related to the halo peak height\footnote{The peak heigh is defined as $\nu \equiv \delta_c / \sigma_R$ where $\delta_c = 1.686$ is the threshold for density collapse, and $\sigma_R$ is the standard deviation of density fluctuations smoothed on the halo Lagrangian scale, $R$.}). The impact of $n_s$ is significantly stronger, particularly for smaller halos. The shape of peaks in the density field---where such peaks eventually collapse, virialize, and form halos---is known to depend directly on the slope of the power spectrum \citep{Bardeen:1986:peaks,Dalal:2010:haloprofiles}. The effective slope of the power spectrum has also been used as one of the inputs in the phenomological model for the concentration--mass relation as presented in \citet{Diemer2019concentrations}. In practice, $n_s$ has a strong impact on the merger history of halos (which is known to impact halo concentrations) as it changes the relative abundance between smaller and bigger halos; the former eventually merges with the latter, and the rate of mergers alter the density profile of the halo.

The impact of $\Omega_{\rm b}$ on the halo concentration is significantly smaller relative to that of the other cosmological parameters, and is always negative; increasing the baryon fraction always reduces the concentration. The smaller amplitude is expected as, unlike the previous cosmological parameters, $\Omega_{\rm b}$ controls a subdominant part of the matter distribution. A 10\% change to the baryonic matter component is only a $0.1 \times 0.15 = 1.5\%$ change to the total matter distribution. The mass dependence of this curve also closely follows the behavior detailed in Figure 7 of \citetalias{\Anba} --- the reduction in halo concentration due to baryons is minimal at Milky Way (MW) mass scales, while there is a larger impact at group scales.\footnote{\citetalias{\Anba} found the concentration \textit{increases} at Milky-Way scales due to the dominance of the cuspy stellar distribution in such halos. However, star formation in the TNG model is resolution dependent \citep{Pillepich2018FirstGalaxies} and at the \textsc{Camels} resolution, we expect fewer stars being formed relative to the results of \citetalias{\Anba} who used the highest-resolution TNG runs.} Note that, even though there is a significant dependence of the halo concentration on these cosmological parameters, this dependence has only small effects on data analyses. This is because we know the cosmology parameters to $\approx 1\%$ precision or better \citep[\eg][]{Planck2020CosmoParams,AbdulKarim:2025:DESIDR2BAO,Abbott:2024:DESSN5YR,DES2022Y3,Asgari2021KidsWL,Anbajagane2025d,Anbajagane2025e}. Thus, the resulting change to the concentration is $\alpha \times 1\% \lesssim 0.8\%$ at all mass scales, given $\alpha = 0.8$ is the largest slope exhibited by a cosmology parameter. For this reason, the cosmology parameters contribute negligible uncertainty to the concentration--mass prediction. Instead, this uncertainty is dominated by the astrophysical parameters as we will note below.

\subsection{Dependence on SNe astrophysics} \label{sec:results:SNeastro}

Moving onto the dependence of halo concentration on different astrophysical parameters, we find a variety of behaviors as expected from the rich phenomology of astrophysical process noted previously. We first discuss the dependence on SNe-related parameters, before moving onto the AGN-related parameters further below in Section \ref{sec:results:AGNastro}.

We start with some general expectations for SNe-related effects on $\cvir$. Following \citetalias{\Anba} (see their Figure 7), we expect SNe feedback to be especially important for halos at MW-mass scales or below, where stellar feedback can efficiently regulate star formation and drive galactic outflows \citep{Dekel1986DwarfGalaxies,Hopkins2014FIRE,Muratov2015GasOutflowsFIRE}. There is also the possibility of non-linear coupling between feedback processes \citepalias{\Shao}, whereby SNe feedback alters the supply and distribution of gas available for black-hole accretion and, consequently, the efficiency with which AGN feedback ejects material beyond the halo radius. All of the SNe parameters considered here affect either the energetics of galactic winds, the star-formation rate and stellar mass of the halo, or its metal distribution \citep{Pillepich2018Methods}. The stellar mass is important in our analysis because the corresponding radial distribution is centrally concentrated thereby increases the measured $\cvir$ \citep{Blumenthal1986AdiabaticContraction,Gnedin2004AdiabaticContraction,Duffy2010BaryonDmProfileDensity,Schaller2015ConcentrationBaryonImprints}. 
There is also non-linear coupling between stellar-mass assembly and SNe feedback, because stronger feedback suppresses and self-regulates star formation \citep{Hopkins2014FIRE}. It is \textit{a priori} difficult to predict the relative importance of enhanced central stellar mass and feedback-driven matter redistribution for a given halo property. The simulated datasets considered here, which evolve the full non-linear connection between these processes,  can reveal which effect dominates the evolution of $\cvir$. Finally, metallicity is an important determinant of the radiative cooling rate of halo gas \citep{Wiersma2009MetalCooling,Pillepich2018Methods}, with radiative cooling being the main way the matter in the halo can lose momentum/energy. Having laid out these general expectations, we now discuss the parameter dependencies in detail.

First, \param{5} is the most dominant astrophysical parameter. It was previously referred to as $A_{\rm SN,2}$ in the \textsc{Camels} release used in \citetalias{\Shao} and scales the SNe wind speed/feedback. As a result, its impact is dominant at/around MW mass scales, and increasing the wind speed always reduces the concentration. \citetalias{\Shao} also found this parameter can \textit{increase} the concentration at high mass, as strong SNe winds disperse the gas distribution away from the halo center, i.e., away from the AGN. As a result, the AGN feedback has less gas to couple to in its vicinity and cannot efficiently eject to distances beyond the halo radius \citep{Gebhardt2023CamelsAGNSN}. Figure \ref{fig:slopes} does not explicitly show this behavior as its uncertainties at $\Mvir > 10^{14}\msun/h$ are too large to identify the behavior.\footnote{This work fits a 28-parameter model to 2048 simulations, while \citetalias{\Shao} fit 6 parameters to 1000 simulations. We incur larger uncertainties per parameter as a result.} The related parameter, \param{3} controls the energy scale of a single SNe explosion, and increasing it both reduces the concentration at low masses and increases it at higher masses. The latter is due to the same AGN coupling effect mentioned previously, but can be observed this time as the impact starts at lower mass-scales (as was also observed in \citetalias{\Shao}) where our uncertainties are significant lower than they are at higher masses.

The next largest impact comes from \param{12}, which was not varied in the older \textsc{Camels} simulation suite used in \citetalias{\Shao}. The Initial Mass Function (IMF) determines the initial population of stars contained within a given star-forming gas cell, and more importantly, the number of SNe events per unit star formation \citep[][see their Equation 11]{Vogelsberger:2013:TNGModel}. The latter term scales the SNe feedback per star-forming cell. As a result, its impact on $\cvir$ follows the same mass-dependent behavior as that of the the \param{3} parameter discussed above (which, like \param{12}, controls the total energy in SNe feedback). The amplitude of the slope is larger as the \param{12} is a power-law index in the integrand that determines the number of SNe while \param{3} is just a linear amplitude coefficient.

The \param{17} raises the floor value of the galactic wind velocity. This velocity scales with the local velocity dispersion of the dark matter surrounding the SNe event, but has a minimum value set by \param{17} \citep[][see their Equation 1]{Pillepich2018Methods}. For sufficiently massive halos, the former will always be larger than the latter and render the value of \param{17} inconsequential. Conversely, for smaller halos, the parameter will have a significant impact. This is precisely what we see in Figure \ref{fig:slopes}, where the dominant impact emerges below $\Mtwohc < 10^{12} \msun/h$. Moving to \param{10} in the upper right panel, increasing its value reduces the efficiency of star formation \citep{Springel:2003:Cosmo}, which in turn reduces the stellar mass fraction and suppresses the amplitude of its centrally peaked matter distribution. While increasing this parameter will also reduce the rate of SNe events, the reduction in the stellar mass fraction causes a bigger effect on $\cvir$. Next, \param{14} controls the fraction of wind energy that is kept as thermal energy instead of kinetic energy, where the former controls the temperature of the gas in the wind itself. Hotter winds stifle star formation in the halo \citep{Pillepich2018Methods} and therefore result in the same effect discussed above, where a less-dominant central stellar component results in a reduced halo concentration.

\param{15} effectively controls the ``mass-loading factor'', which determines how much matter is transported via the winds. The wind velocity is directly tuned by the previously discussed parameters, and so changing the wind momentum changes the mass contained in those winds \citep{Vogelsberger:2013:TNGModel}. Larger values of \param{15} enable more matter to be moved via winds, therefore affecting more of the halo matter distribution (and promoting more metallicity redistribution within the halo). The effect of this parameter is generally small, as it has a slope $|\hat{\alpha}| \leq 0.005$. 

Next, the TNG model also establishes multiple links between SNe winds and the local metallicity. The wind energy is modulated by smoothly suppressing the energy above a given metallicity value of the star-forming cell \citep[][see their Equation 3]{Pillepich2018Methods}. The transition's sharpness in metallicity is set by \param{20}, the metallicity threshold is set by \param{19}, and the amplitude of the reduction is set by \param{18}. The slope with the transition parameter, \param{20}, is essentially zero. There is some increase above $M > 10^{14} \msun/h$ but the slopes are consistent with zero within 2$\sigma$ to 3$\sigma$, so we do not interpret those further. Increases to \param{19} have no impact above group-scale halo masses, but are larger below that scale. Increases to this parameter mean the star-forming cells have to be far more metal rich before their galactic winds get suppressed. The cells become less likely to meet this condition as we increase the threshold and as a result will expel galactic winds with the higher, unsuppressed energies. Naturally, this reduces the halo concentration. The parameter \param{21} controls the fraction of metals deposited immediately into the star-forming cell's vicinity, while the rest are carried away by the winds. Higher values of this fraction result is more inhomogeneous dispersion of metals within the halo. This will make the average gas cell less metal enriched and so, galactic wind energy within the average star-forming cell is not as suppressed by the metallicity-based reduction factor. The increased winds then reduce the halo concentration.

The remaining set of SNe parameters are unique in having \textit{positive} slopes with the halo concentration. \param{16} has the largest amplitude as it scales the gas density below which galactic winds re-couple to the gas cells and evolve hydrodynamically once again. If the density ceiling is increased, the galactic winds only move a short distance before finding an appropriately dense gas cell and recoupling. Thus, the matter in the winds are not dispersed as efficiently through the halo, and results in a more concentrated halo. \param{11} controls the effective equation-of-state (EoS) of the gas, i.e. the gas pressure. Larger values increase this pressure and prevent the gas from forming high-density regions that eventually produce star formation. As a result, high pressure results in fewer SNe events and therefore increases the halo concentration. Finally, \param{13} sets the minimum mass used in the mass integral that sets the SNe rate; the same mass integral discussed for the \param{12} \citep[][see their Equation 11]{Vogelsberger:2013:TNGModel}. Higher values of \param{13} narrow the range of the integral and therefore reduce the SNe rate, which in turn increases the halo concentration.

\subsection{Dependence on AGN astrophysics} \label{sec:results:AGNastro}

We now move onto the AGN feedback parameters. The AGN feedback in TNG is split into two phases --- (i) a thermal/quasar phase, which dominates the Milky-Way mass and above, and (ii) the kinetic/radio phase, which dominates at group-scale masses and above \citep{Weinberger2017Methods}. The two phases are also referred to as high- and low-accretion phases, given the AGN follows each regime when its mass accretion is high/low enough relative to the Eddington rate \citep{Eddington1916RadiativeEquilibrium}, which sets the maximum luminosity (i.e. radiation pressure) that can be supported by a given gravitational field. The thermal phase rapidly injects heat into the gas in the vicinity of the AGN, while the radio phase generates jets that inject kinetic energy and normally expel the gas beyond the halo radii, $\Rvir$.

Focusing on the original two AGN parameters present in the first \textsc{Camels} release, \param{4} and \param{6}, we see the expected phenomenology discussed in \citetalias{\Shao}. \param{4} changes the energy scale per kinetic feedback event, and increasing it naturally reduces the halo concentration. As was noted in \citetalias{\Shao}, the impact here is minimal relative to that of the other parameters discussed thus far: a factor of 2 change in the energy scale alters the concentration by only 2\%. The \param{6} parameter changes the rate of AGN events, which in turn inversely impacts the corresponding energy scale per event (i.e., we either have many lower-energy events or few higher-energy ones). In low-mass halos, numerous low-energy events are less likely to eject material outside the virial radius, and therefore do not reduce the concentration as much. As a result the slope with this parameter is positive for most masses. For higher-mass systems, the AGN is large enough that the low-energy events may still be sufficient to eject material outside the halo radius. In this case, having more events may make the AGN \textit{more} efficient at ejecting material out the halo. However, the slope at these higher masses is still consistent with $\hat{\alpha} = 0$ within $\approx 3\sigma$ so we do not interpret it further at this time. Dedicated analysis of the AGN events, using particle-level data in the simulations, is required to disentangle these features further.

Increasing \param{22} results in stronger feedback and lower concentration, as the AGN starts from a more massive progenitor and can grow faster (black hole mass is tied to its accretion rate; see \citealp{Weinberger2017Methods}), resulting in stronger feedback. The next two parameters scale the accretion rate of the SMBH. This rate is set by the combination of the Bondi rate \citep{Bondi1952Accretion} and the Eddington rate. The former is the mass-accretion scale set by how quickly the gas falls into the SMBH, and depends on both the density and sound speed of gas in the vicinity of the AGN. The Eddington rate is set by the radiation pressure limit as noted above \citep{Eddington1916RadiativeEquilibrium}. In the TNG model, the SMBH accretion rate is set by the smaller of the two rates. \param{23} controls the Bondi accretion rate. Increasing it causes the AGN to more predominantly operate in a thermal feedback state; see Equation 1 and 3 in \citet{Weinberger2018SMBHsIllustrisTNG}. This means it rapidly heats the nearby gas, which in turn transfers the heat further out into the halo. The thermal feedback process is known to reduce the halo concentration at MW mass-scales, as shown in \citetalias{\Anba}. Increasing the Eddington rate, as set by \param{24}, will put the halo into a predominantly kinetic feedback state. This appears to have only a minor effect on the halo concentration, largely mimicking the other radio feedback parameter we discussed thus far, \param{4}.

Next, \param{27} and \param{28} control the normalization and power-law scaling with mass, respectively, of the threshold (defined in the Bondi-to-Eddington ratio) below which the AGN operates in thermal mode. Increases to \param{27} results in AGN of all host halo masses more frequently operating in thermal feedback mode. As a result, the concentration is reduced at all halo masses. Increases to \param{28} do the same for more-massive halos while making it more challenging for less-massive halos to fall below this threshold. As a result, the slopes in Figure \ref{fig:slopes} show a zero crossing where less massive halos have their concentration increased instead. Finally, the two remaining parameters, \param{25} and \param{26}, control the energy efficiency and the radiative efficiency, respectively. While both parameters scale the energy rate of thermal feedback, the latter also scales the Eddington accretion rate \citep[][see their Equation 3]{Weinberger2017Methods}. Changing the Eddington rate will naturally also alter whether the halo is in its thermal or kinetic feedback state. Interestingly, increases to these two parameters result in more concentrated halos. This is because a more efficient thermal feedback phase heats the gas surrounding the AGN and stifles further accretion and AGN growth \citep{Booth2009AGNFeedback,Weinberger2017Methods}. As a result, the AGN does not achieve its high-mass, low-accretion, kinetic mode \citep{Weinberger2018SMBHsIllustrisTNG,Terrazas2020SMBHsInTNG}. The stronger dependence on \param{26} (compared to \param{25}) is also expected as this parameter directly reduces the AGN growth by reducing the Eddington accretion ratio \citep[][see their Equations 3, 7, and 9]{Weinberger2017Methods}. \param{25} on the other hand has only a secondary effect as it controls the properties of the AGN's environment --- as it controls the heat emitted by the AGN emits into the nearby gas --- rather than the AGN growth itself \citep{Booth2009AGNFeedback,Weinberger2017Methods}.

From the discussion above, it may seem counter intuitive that the impact of AGN on the halo concentration is largest when changing the thermal feedback behavior rather than the kinetic feedback given the latter is the more energetic of the two. However, \citetalias{\Anba} (see their Figure 7) show that for the cluster-scale halos where the kinetic feedback energy peaks, the gas cooling energy rate is still higher than the kinetic feedback energy rate. \citetalias{\Anba} also show that for such halo masses, the halo concentration in TNG agrees closely with the dark matter-only variant.  Given this result, changes to the AGN kinetic feedback behavior may have less of an impact on the concentration than expected. Thus, we treat the de-emphasis of jet-related parameters to be a consequence of the TNG model behavior and find it is consistent with existing, independent analyses of the original TNG simulations.

\subsection{Principle components}\label{sec:results:PCA}

\begin{figure}
    \centering
    \includegraphics[width=\columnwidth]{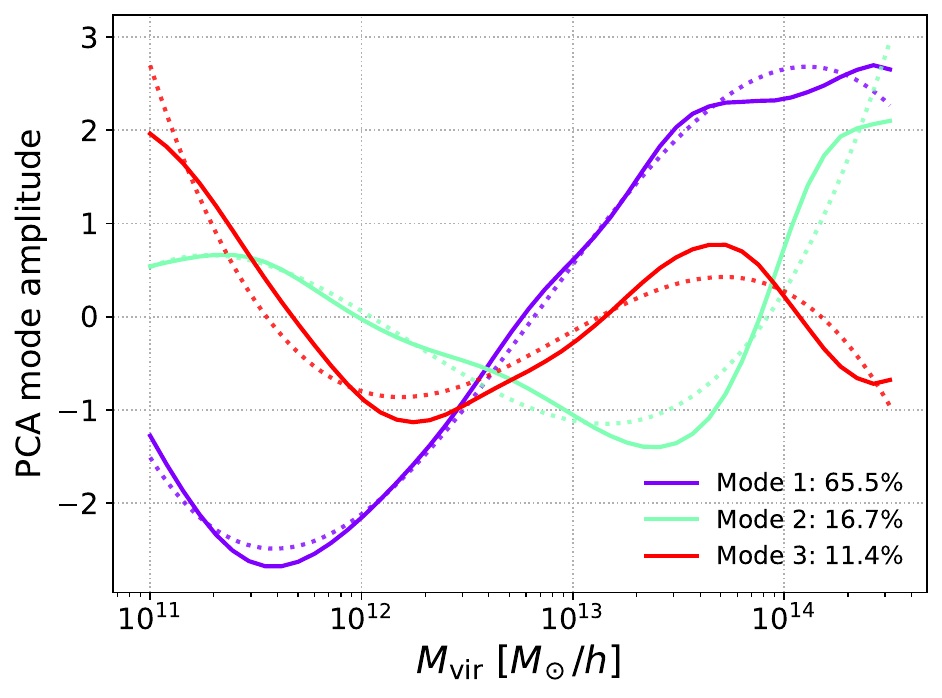}
    \caption{The mass dependence of the three components from our PCA approach. The solid lines show the components while the dotted lines are our cubic approximation (Equation \ref{eqn:cubic}).}
    \label{fig:PCA_modes}
\end{figure}

Our analysis has produced the mass-dependent derivative $\hat{\alpha} = d \ln \cvir(\Mvir) / d\ln X$ for 23 astrophysical parameters, $X$. As discussed in Sections \ref{sec:results:SNeastro} and \ref{sec:results:AGNastro}, the shape of the mass-dependence in the slopes is often similar across parameters given they impact a common set of SNe/AGN processes, and the nature of the processes set the main mass dependence. As we have a large number of slope estimates for astrophysical parameters (23 parameters relative to the 4 extracted in \citetalias{\Shao}) we can decompose these trends into a low-order principle component basis. We find that three components are sufficient to capture roughly 95\% of the variations across the different slopes. These components are displayed in Figure \ref{fig:PCA_modes} and are characterized by the parameter combinations,
{
\allowdisplaybreaks[4]
\begin{align}\label{eqn:PCA_def}
    A_1 ={}&\hspace{10pt}0.368\,\,\param{5} \nonumber\\
           &+ 0.346\,\,\param{3} \nonumber\\
           &+ 0.345\,\,\param{12} \nonumber\\
           &+ 0.303\,\,\param{14}\\[10pt]
    A_2 ={}&\hspace{10pt}0.446\,\,\param{27} \nonumber\\
           &+ 0.414\,\,\param{6} \nonumber\\
           &+ 0.349\,\,\param{20} \nonumber\\
           &+ 0.344\,\,\param{28}\\[10pt]
    A_3 ={}&\hspace{10pt}0.532\,\,\param{23} \nonumber\\
           &- 0.409\,\,\param{25} \nonumber\\
           &- 0.306\,\,\param{17} \nonumber\\
           &+ 0.289\,\,\param{21}
\end{align}
}
where each term on the right-hand side should be viewed as the corresponding mass-dependent slopes of that parameter. To construct this basis, we normalize the input slopes such that the maximum of their absolute value is unity. The first mode is dominated by SNe parameters, while the second and third are dominated by the AGN parameters (though some SNe parameters are present in the latter two as well). Equation \eqref{eqn:PCA_def} only lists the four most-dominant parameters defining each component.

We find that these components can be approximated by a simple cubic polynomial in log mass,
\begin{align}\label{eqn:cubic}
    \mathcal{P}_{1}(\Mvir) &= -0.2 & + \,\,&  3.1\,\mu &+\,\,& 0.19\,\mu^{2} &-\,\,& 0.65\,\mu^{3} \\
    \mathcal{P}_{2}(\Mvir) &= -0.93 & - \,\,&  1.0\,\mu &+\,\,& 0.87\,\mu^{2} &+\,\,& 0.55\,\mu^{3} \\
    \mathcal{P}_{3}(\Mvir) &= -0.46 & + \,\,&  1.2\,\mu &+\,\,& 0.44\,\mu^{2} &-\,\,& 0.73\,\mu^{3}
\end{align}
where we have defined, $\mu(\Mvir) \equiv \log_{10}(\Mvir/M_{\rm piv})$ with $M_{\rm piv} = 5.6 \times 10^{12} \msun/h$. We show these approximations as the dotted lines in Figure \ref{fig:PCA_modes}.

\begin{figure}
    \centering
    \includegraphics[width=\columnwidth]{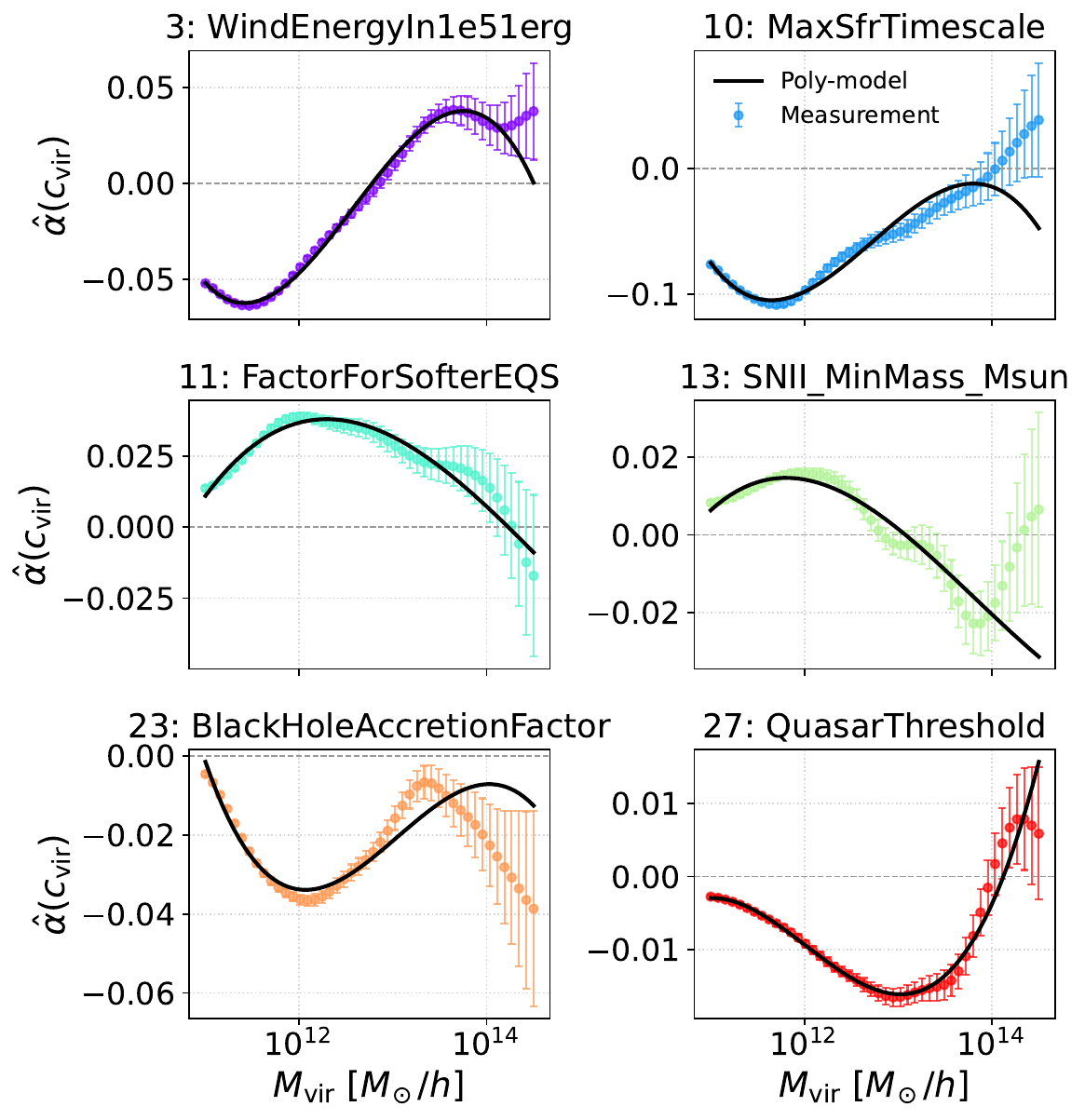}
    \caption{The measured slopes for six different parameters, alongside a fit of the three-component model in Equation \eqref{eqn:cubic}. The errorbars are the same ones shown in Figure \ref{fig:slopes}. The three-component model is able to reproduce the mass-dependent behavior exhibited by the various parameters. While we show the slopes for $z = 0$, a similarly accurate fit is obtained for slopes across $z \in [0, 3]$.}
    \label{fig:PCA_fit}
\end{figure}

These components (and their cubic approximations) are useful for analyses that wish to marginalize over unknown astrophysical effects in the concentration--mass relation, but are not focused on constraining the physical parameters pertaining to those effects. In this case, the three-component model of Figure \ref{fig:PCA_modes} provides a simple, approximate alternative for such a model. While we construct our modes using the slopes at $z = 0$, we have found the three modes can still capture the slopes within $0 < z < 3$ as the higher redshifts have similar mass-dependent shapes to their $z = 0$ counterparts (Section \ref{sec:results:zdep}). Figure \ref{fig:PCA_fit} shows the fit of these modes to a few of the mass-dependent slopes discussed in Figure \ref{fig:slopes}. We find an adequate fit, especially when comparing the residuals to the bootstrap-derived uncertainties on the slope. Thus, the cubic polynomials can adequately represent the effective impact of astrophysics --- as determined by the TNG model --- on the halo concentration.

\subsection{Redshift variations}\label{sec:results:zdep}

\begin{figure*}
    \centering
    \includegraphics[width=2\columnwidth]{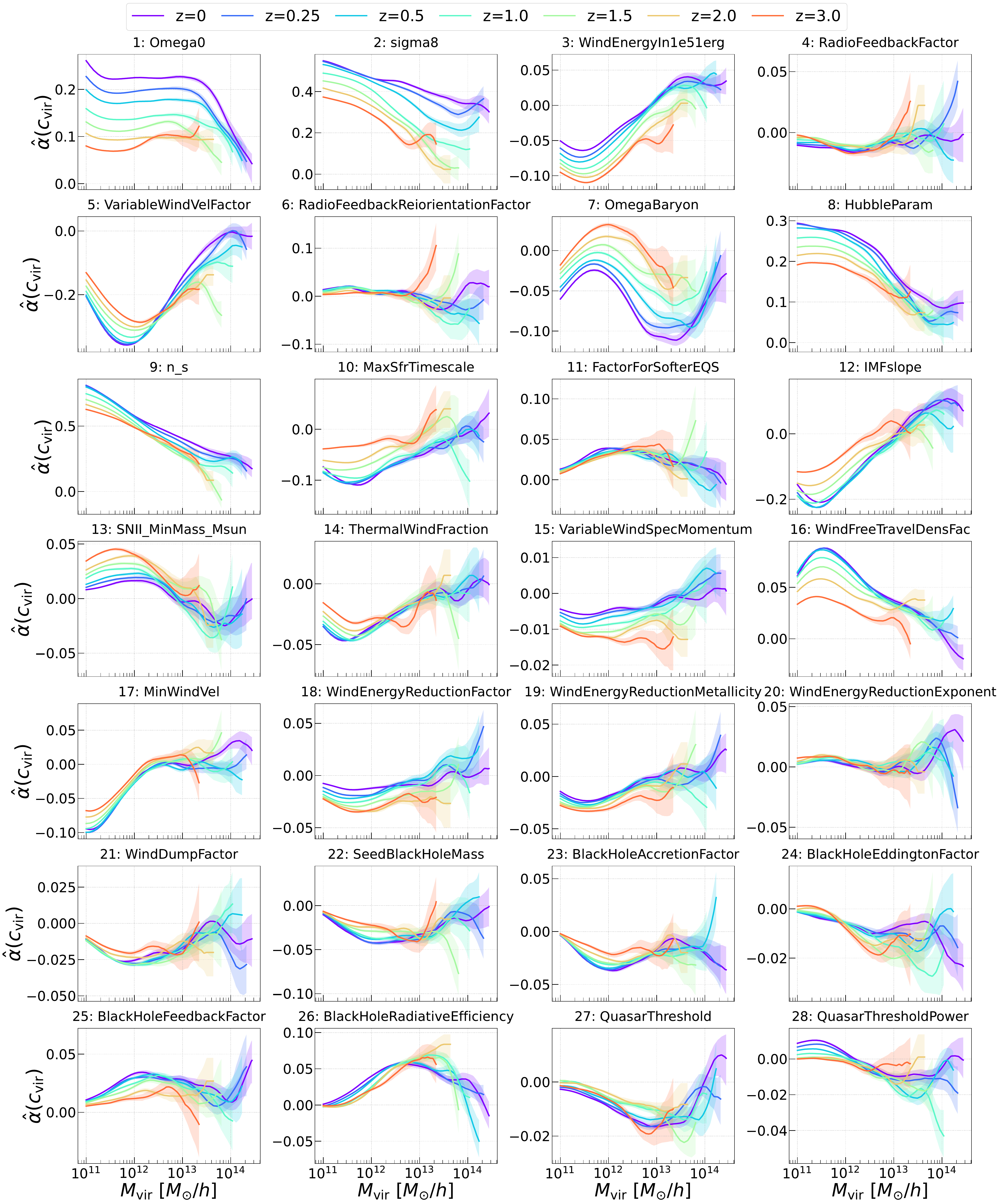}
    \caption{The KLLR slopes between halo concentration $\cvir$ and the twenty-eight astrophysical and cosmological parameters (denoted by text in each panel) at redshifts $0 < z < 3$. The slopes are only plotted up to the 20th most-massive halo in the uber sample, to avoid showing mass ranges with poor statistics.}
    \label{fig:slopes_zdep}
\end{figure*}

Our analysis thus far has been limited to $z = 0$. We now extract the slopes for higher redshifts, and demonstrate that the phenomenology discussed thus far extends to these redshift regimes. Figure \ref{fig:slopes_zdep} presents the key trends.

All five cosmology parameters show a weaker connection with $\cvir$ at higher redshifts. This is also found in the simulation-calibrated model of \citet{Diemer2019concentrations}. This behavior can generally be expected as structures at low redshifts have had more time to be impacted by gravitational evolution, where this evolution is strongly dictated by the cosmological properties (i.e., parameter values) of the simulation \citep{Dodelson2003ModernCosmology}. Thus, objects are low (high) redshifts are more (less) sensitive to cosmology. We note that the shape of the mass-dependence exhibited by the slopes are broadly similar across the different redshifts.

For the slopes with the astrophysical parameters, their amplitudes both increase and decrease with redshift, depending on the exact parameter. No slopes completely change sign across redshift (i.e. no instance of a slope being all positive at one redshift and all negative at another). Prior to discussing the results in detail, we first discuss the general expectations. SNe feedback is more prevalent at higher redshift as the peak of star formation is around $z \approx 2$ \citep{Wechsler2018GalaxyHaloConnection}. For most massive halos, the thermal feedback mode of the AGNs is more energetically important at similarly high redshifts, while the kinetic feedback dominates mostly lower redshifts. \citet{Weinberger2018SMBHsIllustrisTNG} show these specific redshift dependence of the feedback energies in the TNG model. However, this does not mean all SNe and AGN parameters will increase their amplitudes with redshift. Some parameters, such as \param{5}, have a redshift-dependent definition \citep[][see their Equation 1]{Pillepich2018Methods}, which can cause its behavior to deviate from the above expectation.

The parameters with the largest redshift evolution are all SNe parameters. Here, ``large'' is defined as the fractional difference in the slope values when compared to their $z = 0$ values.\footnote{The fraction is not a robust metric as some slopes have zero crossings, where a fraction is ill-defined. However, such slopes do not exhibit dramatic evolution eitherway, so we still use this metric as a rough standard.} The SNe parameters with increasing redshift dependence are \param{3}, \param{13}, \param{15}, \param{18}, \param{19}. These parameters all control the baseline SNe feedback energy value per unit star formation. Therefore they all increase with redshift, as there is more total star formation per halo at higher redshifts. Interestingly, all remaining SNe parameters have a reduced impact at higher redshifts. This includes \param{10}, \param{12}, \param{14}, \param{16}, \param{17}, \param{21}. These parameters control a wide variety of processes connected to SNe, but do not share a single physical role. They instead modify the local star-formation or stellar-population model (\param{10}, \param{12}), the propagation of galactic winds (\param{14}, \param{16}, \param{17}), or the redistribution of metals (\param{21}), rather than acting solely as normalizations of the total feedback budget \citep{Pillepich2018Methods,Ni2023Camels28}. A plausible interpretation of their common decline with redshift is that these effects are more indirect and cumulative: at high redshift, star formation is governed more strongly by rapid halo growth, gas supply, and feedback self-regulation, whereas the detailed star-formation prescription, repeated wind propagation, stellar buildup, and metal enrichment have greater leverage on halo structure at later times \citep{Schaye2010CosmicSFH}. However, because these parameters affect distinct pieces of the subgrid model, we do not interpret their common redshift trend as evidence for a single underlying physical mechanism.

Almost all AGN feedback parameters exhibit a common phenomenology, with their impact decreasing toward higher redshift. For many of these parameters, this behavior may be expected because they control the kinetic feedback mode, whose energy scale decreases with increasing redshift over the range of halo masses considered here \citep{Weinberger2018SMBHsIllustrisTNG}. However, our discussion in Section \ref{sec:results:AGNastro} highlighted the generally weak direct impact of kinetic-mode feedback on $\cvir$. It is therefore more likely that $\cvir$ is sensitive to the accumulated effects of AGN feedback over time. At higher redshift, less feedback energy has been deposited into the halo, and the halo has had less time to dynamically respond to this energy injection. Consequently, variations in the AGN parameters produce a weaker change in $\cvir$. This interpretation is particularly relevant for parameters that control black-hole accretion and growth, such as \param{23} and \param{24}, whose effects naturally accumulate over the growth history of the black hole. A more detailed analysis using the thermodynamic properties of particles stored in the three-dimensional simulation snapshots is required to identify the underlying mechanism of this redshift evolution. We leave such an analysis to future work.

\section{Conclusions}\label{sec:conclusions}

The halo concentration is a ubiquitous tool in modeling both the internal structure of halos and the large-scale features of the cosmic density field. The concentration--mass relation is therefore widely used in various models of this structure. However, the rich variety of astrophysical processes can leave distinct features on this relation. In this work, we utilize a set of simulations from the \textsc{Camels} suite, simulated under the TNG galaxy formation model, to characterize the impact of 28 different astrophysical/cosmological parameters on this relation. We use a kernel-localized regression method that can capture the mass-dependent impact of astrophysics on the halo concentration. Throughout, the concentration is estimated through the \textsc{Rockstar} halo finder and represents the total mass content of the halo and not just the dark matter.

Our main results are as follows:
\begin{itemize}
    \item The cosmology parameters have the largest impact on the concentration and generally have a stronger impact on smaller halos (Figure \ref{fig:slopes}).

    \item The SNe-related parameters all show similar phenomenology of an increasing impact at MW mass-scales, with larger parameter values increasing galactic wind energy/speed and leading to a less concentrated halo. For some SNe parameters, increasing their values leads to higher concentrations as they increase the mass fraction of the centrally peaked stellar core. In all cases, the dominant impact is at or near the MW mass scale.

    \item Following existing analyses of \textsc{Camels}, we find more evidence of the non-linear coupling of SNe and AGN feedback. Increasing SNe feedback evacuates gas near the AGN, and leads to less gas in the AGN vicinity that can be ejected via AGN feedback. Thus, at higher halo masses, where AGN feedback is more dominant, the increase of SNe feedback increases the halo concentration.

    \item The largest slope with an AGN parameter has a significantly smaller amplitude than its counterpart for the SNe case. However, many AGN parameters still have slopes of comparable amplitude to those of various SNe parameters. The dominant impact of the AGN is at higher-mass scales than the SNe and the parameters both increase or decrease the halo concentration.

    \item All slopes show largely similar behavior across redshift. The amplitude of the slopes for the SNe parameters increase or decrease with redshift depending on the parameter definition. The slope amplitudes for the AGN and cosmology parameters always decrease with redshift (Figure \ref{fig:slopes_zdep}).
    
    \item We build a simple 3-component polynomial basis that can capture the various mass-dependent impacts of astrophysics on the halo concentration (Figure \ref{fig:PCA_modes}). These modes are sufficient to fit the slopes for a given parameter (Figure \ref{fig:PCA_fit}), and does so across $0 < z < 3$.
\end{itemize}

The halo concentration traces the evolution and interaction of different astrophysical components, and of the processes governing them. The astrophysical imprints are not emergent only with a narrow mass range and are instead generically present across all halo masses studied in this work. Halo-model analyses that focus on cosmic structure at the halo scale or smaller will need to consider such effects in their modeling. One approach is to fully model the individual components of the halo \citep[\eg][]{Schneider2019Baryonification, Arico:2020:Baryons, Anbajagane:2024:Baryonification}, and marginalize over uncertainties in these components. A cruder (but simpler and quicker) approach to retain the NFW prescription for the profile but modulate the halo concentration to mimic the impact of baryons. In such cases, the results presented in our work will be a useful reference point. The three-component model we introduce enables rapid marginalization of astrophysical effects on the concentration--mass relation, instead of having to consider each of the 23 astrophysical parameters discussed in our work.

All analysis and interpretation discussed here has been limited to the IllustrisTNG model of galaxy formation. The integrated baryon properties of halos are known to vary significantly across different galaxy formation models \citep[\eg][]{Farahi2018BAHAMAS, Chua2019ShapeIllustrisBaryons, Anbajagane2020StellarStatistics, Chua2021TNGShapes, Lim2021GasMass, Anbajagane2022BaryImprint, Anbajagane2022GalVelBias, Stiskalek2022ML}, and so the concentration--mass relation naturally will exhibit such a difference as well. The \textsc{Camels} suite continues extending their 28-parameter approach to other galaxy formation models, providing an avenue to quantify these effects. Future work will be able to more generally check the reliability of our component-based model in capturing the astrophysical imprints on the halo concentration across the different simulations.

\section*{Acknowledgements}
We thank the \textsc{Camels} collaboration for making their data publicly available and easily accessible to the community. CD was supported by the Metcalf Summer Experience Grant. DA is supported by NSF grant No. 2108168. 

All analysis in this work was enabled greatly by the following software: \textsc{Pandas} \citep{Mckinney2011pandas}, \textsc{NumPy} \citep{vanderWalt2011Numpy}, \textsc{SciPy} \citep{Virtanen2020Scipy}, and \textsc{Matplotlib} \citep{Hunter2007Matplotlib}. We have also used
the Astrophysics Data Service (\href{https://ui.adsabs.harvard.edu/}{ADS}) and \href{https://arxiv.org/}{\texttt{arXiv}} preprint repository extensively during this project and the writing of the paper.

\section*{Data Availability}

The public \textsc{Camels} data, including the \textsc{Rockstar} catalogs we use in this work, can be found at \url{https://camels.readthedocs.io}. The \textsc{kllr} method we use is publicly available at \href{https://github.com/afarahi/KLLR}{github.com/afarahi/KLLR}.

\bibliographystyle{mnras}
\bibliography{References}

\appendix

\label{lastpage}
\end{document}